\documentclass[aps,prl,showpacs,twocolumn,amsmath,amssymb,superscriptaddress]{revtex4-1}

\usepackage{tabularx}
\usepackage{bm}
\usepackage{graphicx}

\usepackage{color}

\usepackage{multirow}
\usepackage{dcolumn}
\usepackage{amssymb,amscd,xypic,bm,wasysym}
\usepackage{float}
\usepackage{cleveref}
\newcolumntype{Y}{>{\centering\arraybackslash}X}

\usepackage{soul}  

\begin{document}

\title{Quantum Anomalous Hall Effect through Canted Antiferromagnetism}

\author{Xiao Li}
\affiliation{Center for Quantum Transport and Thermal Energy Science, School of Physics and Technology, Nanjing Normal University, Nanjing 210023, China}
\author{Allan H. MacDonald}
\affiliation{Department of Physics, The University of Texas at Austin, Austin, TX 78712, USA}
\author{Hua Chen}
\affiliation{Department of Physics, Colorado State University, Fort Collins, CO 80523, USA}
\affiliation{School of Advanced Materials Discovery, Colorado State University, Fort Collins, CO 80523, USA}
    
\begin{abstract}
Most canted antiferromagnets are also anomalous Hall antiferromagnets (AHE AFMs), {\it i.e.} they have an anomalous Hall response and other responses with the same symmetry requirements. We suggest that AHE AFMs are promising materials as hosts for high-temperature quantum anomalous Hall effects. By considering models of two-dimensional (001) perovskite layers with strong spin-orbit coupling that isolates an effective total angular momentum $\tilde{j}= \frac{1}{2}$ subspace of the $t_{2g}$ manifold, we propose a strategy to engineering quantum anomalous Hall antiferromagnets.
\end{abstract}

\maketitle

\textit{Introduction---}
The anomalous Hall effect (AHE) refers to transverse charge current response to electric fields that changes sign under time reversal, but does not require external magnetic fields. It is now well established that the anomalous Hall conductivity, which can be viewed as a time-reversal-odd pseudovector \cite{chen_2014, chen_2018}, is allowed by symmetry and relatively large in many noncollinear antiferromagnets (AFMs) \cite{chen_2014, kubler_2014, nakatsuji_2015, nayak_2016}. Whenever the anomalous Hall effect is present other physical observables which can, like the Hall conductivity, be viewed as time-reversal-odd pseudovectors, such as the Kerr \cite{feng_2015,higo_2018}, Faraday (with caveats \cite{armitage_2014}), and anomalous Nernst effects \cite{ikhlas_2017}, are also present and remain large even if the total magnetization accidentally vanishes.

We have previously proposed referring to antiferromagnets with large AHEs as AHE AFMs \cite{chen_2018}. Since the net magnetization, which has both spin and orbital contributions\cite{chen_2018}, is generically non-zero but typically very small in AHE AFMs, these materials are also often referred to as weak ferromagnets. The AHE AFM terminology emphasizes the important consequences of symmetry \cite{dzyaloshinskii_1958} in these magnetic structures, which have electrical and optical properties that are more similar to those of regular strong ferromagnets. The study of weak ferromagnetism has a long history, and has identified many examples, including NiF$_2$\cite{moriya_1960_1}, $\alpha-$Fe$_2$O$_3$ \cite{dzyaloshinskii_1958, moriya_1960_2}, and orthoferrites \cite{bozorth_1958, geller_1956}, etc. Most have nearly collinear antiferromagnetic order. The most common origin of weak magnetization is spin-orbit-interaction driven canting of local moment directions relative to those in the ideal collinear AFM structure.

The quantum anomalous Hall effect (QAHE) is an extreme form of the AHE which can appear in two dimensional (2D) insulators. In the QAHE, a quantized Hall conductance is carried by dissipationless chiral edge channels which are protected by bulk Bloch bands with nonzero Chern numbers \cite{haldane_1988, chang_2013}. There have been many theoretical proposals for realizations of the QAHE, most \cite{liu_2008, yu_2010, qiao_2010, tse_2011, jiang_2012, jiang_2012_2, qiao_2012, zhang_2012, wang_2013, qiao_2014, cai_2015}, if not all \cite{liang_2013, zhang_2014, ezawa_2015, zhou_2016}, with time-reversal symmetry broken by ferromagnetism. The only experimentally established system, magnetically doped topological insulators, has time-reversal broken by delicate ferromagnetic ordering of the magnetic dopants, and as a consequence requires very low temperatures for observation of the quantized Hall conductance \cite{chang_2013}. The most common strategy for identifying more robust QAHE systems is to look for semi-metals with very small Fermi surfaces. Formation of a QAHE state then requires only that these small regions of the Brillouin zone be gapped \cite{haldane_1988} by appropriate time-reversal symmetry breaking mass terms. Because itinerant electron antiferromagnetism has the general tendency of reducing the density of states at the Fermi energy even in robust metals with large Fermi surfaces, we propose that the potential for achieving more robust QAHE states is greater in AHE AFMs than in itinerant electron ferromagnets which are rarely insulators. In this Letter we study a model QAHE system based on canted antiferromagnetism on a 2D square lattice, and discuss its possible realization in orthorhombic $4d$ or $5d$ perovskite (001) thin films.

\textit{AHE in bulk collinear AFMs---}
We start by giving some concrete model examples of how symmetry-allowed canting can lead to an AHE in collinear AFMs.  The two classic examples of weak ferromagnetism are--NiF$_2$ \cite{moriya_1960_1} and $\alpha-$Fe$_2$O$_3$ (hematite) \cite{dzyaloshinskii_1958,moriya_1960_2}. When they are described by spin models, canting can be driven either by site-dependent single-ion anisotropy or by anisotropic exchange interactions. Although the classical weak ferromagnets 
are very good insulators, and not ideal for observing an AHE even when doped \cite{zhao_2011}, their properties are still instructive. In particular the properties of NiF$_2$ are related to the QAHE models that we introduce below. Toy models motivated by the properties of hematite are discussed in the supporting material \cite{supp}. 

\begin{figure}
	\begin{center}
		\includegraphics[width=2 in]{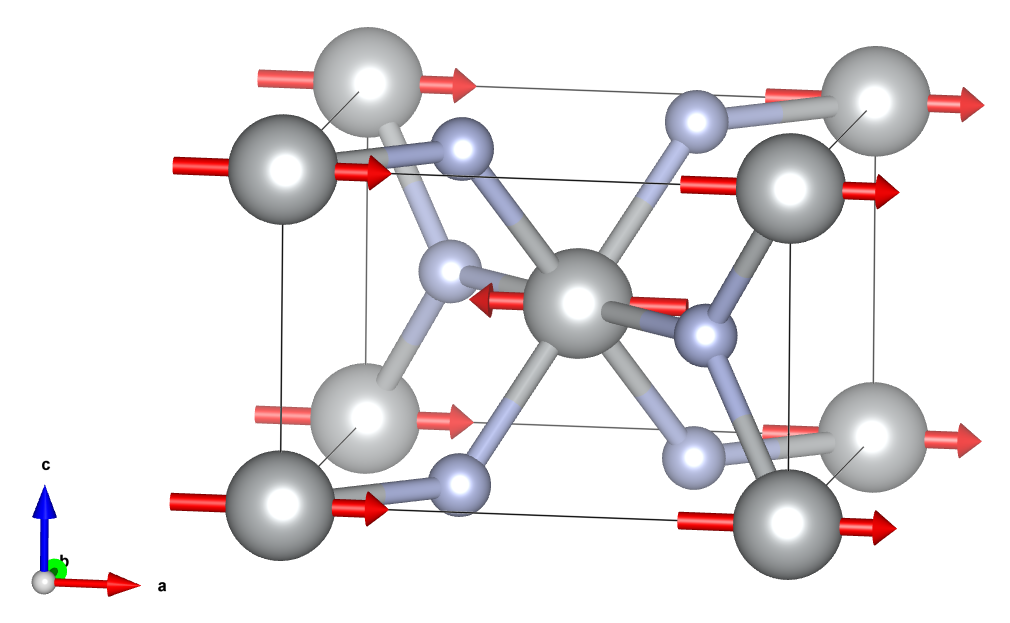}
	\end{center}
	\caption{(Color online) Structure and magnetic order of NiF$_2$. The large and small balls represent Ni and F atoms, respectively. 
	The arrows indicate the directions of the local magnetic moments on the Ni atoms.}\label{fig:1}
\end{figure}

NiF$_2$ (Fig.~\ref{fig:1}) has a rutile structure with Ni atoms on a distorted bcc lattice in which the lattice constant along $c$ is shorter than equal lattice constants along $a$ and $b$. The unit cell has two Ni atoms, each of which is sitting at the center of an octahedron formed by 6 F atoms, but the two octahedra differ by a $\pi/2$ relative rotation around $c$. The site anisotropy of the Ni moments in the $ab$ plane, defined by the F-Ni-F bonds, is therefore different between the two Ni sites and favors a $\pi/2$ difference in orientation that is in conflict with perfect collinear antiferromagnetism. When both moments lie in the $ab$ plane and couple to each other antiferromagnetically, anisotropy will align the N\'{e}el vector along $a$ (or $b$) with weak canting and a net moment along $b$ (or $a$) \cite{moriya_1960_1}.  

Systems in which time-reversal accompanied by a translation is a symmetry cannot support an AHE. Since this symmetry is absent in NiF$_2$ even when the moments are forced to be perfectly collinear, we expect systems with the NiF$_2$ symmetry to be AHE AFMs. We have constructed a minimal $s-d$ model for NiF$_2$-like materials by assuming $s$-electron hopping on the bcc lattice formed by the Ni atoms, with a symmetry-allowed spin-orbit coupling term constructed using the recipe given in \cite{chen_2018}. Specifically, since inversion symmetry with respect to the center of a Ni-Ni bond along a main diagonal of the cubic unit cell is broken by the off-center F atom, the spin-dependent hopping along such a bond should resemble the microscopic spin-orbit interaction [$\propto (\nabla V \times \bm p)\cdot \bm s$], by changing sign with hopping direction and being proportional to $\bm s \cdot \bm n$ where $\bm n$ is a vector that is perpendicular to the plane containing the Ni and F atoms. A staggered local Zeeman field is added to mimic the antiferromagnetic order in Fig.~\ref{fig:1}. 

We calculated the intrinsic anomalous Hall conductivity vector $\bm \sigma_{\rm AH}$ and the orbital magnetization $\bm M_{\rm orb}$ of this 4-band model using different sets of parameter values, and found that they are generically nonzero. 
The energy is lowest when the N\'{e}el vector is along the $a$ or $b$ axes, in which case both $\bm \sigma_{\rm AH}$ and $\bm M_{\rm orb}$ are perpendicular to the N\'{e}el vector and in the $ab$ plane, as expected from symmetry considerations. When the N\'{e}el vector deviates from the high symmetry directions, the lowered symmetry allows $\bm \sigma_{\rm AH}$, $\bm M_{\rm orb}$, and the net spin magnetization $\bm M_{\rm spin}$ to point in different directions. These effects must be considered when studying coherent reorientation of the order parameter using a magnetic field \cite{chen_2018}, for which we presented a detailed analysis in \cite{supp}.

In hematite, Dzyaloshinskii-Moriya interactions dominate over single-ion anisotropy because trigonal symmetry along the [111] direction eliminates uniaxial single-ion anisotropy of Fe moments in the (111) plane. The same symmetry consideration complicates construction of the essential spin-orbit coupling term in the corresponding minimal model for hematite structure antiferromagnets \cite{supp}. However, the two mechanisms do not necessarily correspond to qualitatively different behaviors of the AHE. Although it is not as commonly mentioned as the above two mechanisms, a two-site anisotropy, or anisotropy in the symmetric part of the exchange coupling tensor between local moments, can also lead to canting \cite{szunyogh_2009}.

\begin{figure}
	\begin{center}
		\includegraphics[width=1.7 in]{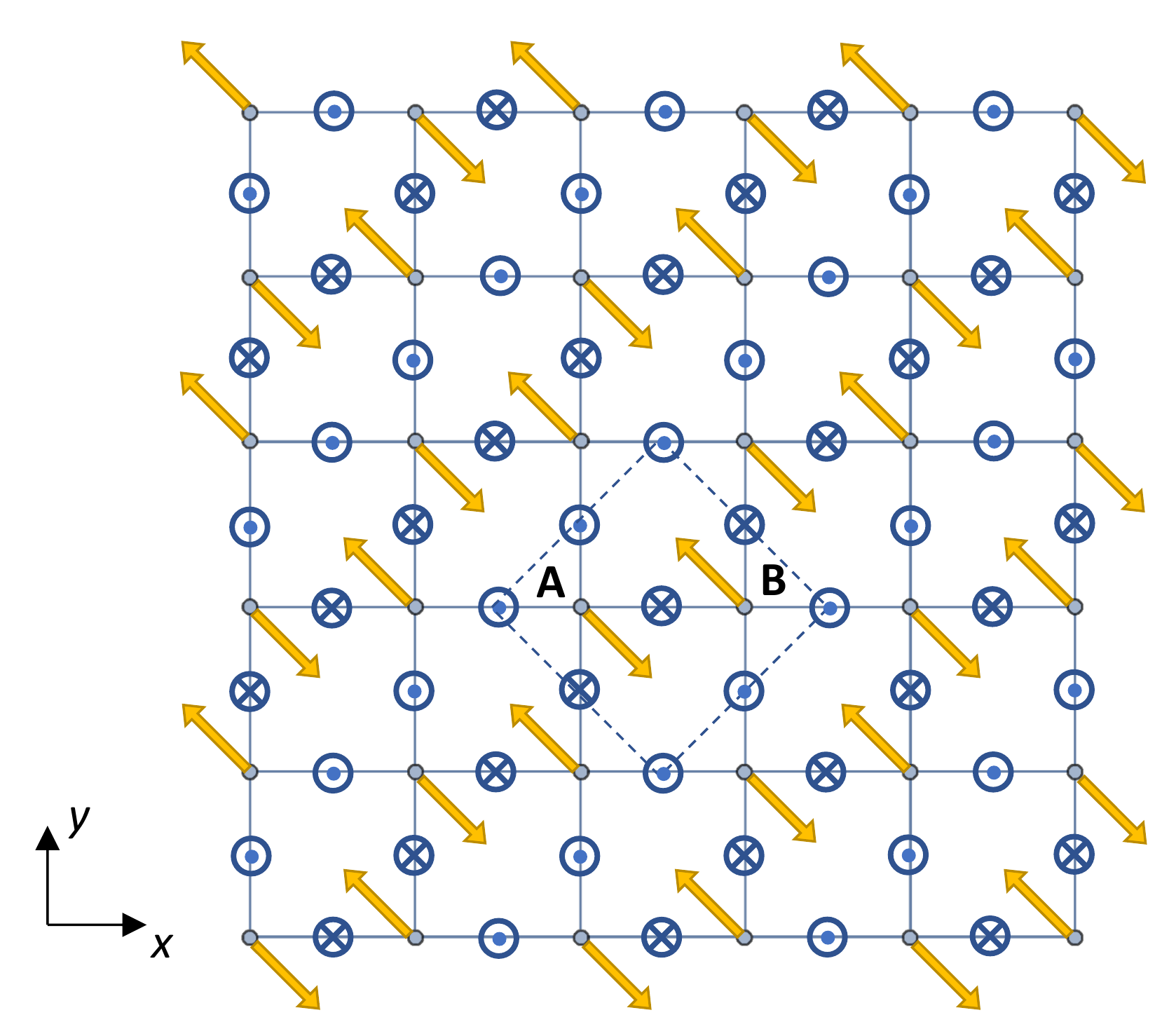}
	\end{center}
	\caption{(Color online) A 2D model of a collinear AFM on a square lattice that can give rise to QAHE. The orange in-plane arrows on the 2D lattice sites represent the exchange fields associated with magnetic order while the out of plane arrows on the 2D lattice links represent the direction $\hat{\eta}$ in the nearest-neighbor hopping spin-orbit coupling terms (Eq.~\ref{eq:Hnif22d}).} \label{fig:2}
\end{figure}

\textit{QAHE in 2D Canted AFMs: Building a Toy Model---}
We next propose a two-dimensional (2D) square lattice model based on the NiF$_2$ structure, and show that it can host a QAHE supported by canted AFM order. The model, summarized in Fig.~\ref{fig:2}, is motivated by (011) planes of NiF$_2$, which have the same two-sublattice unit cell as the bulk structure, an in-plane N\'{e}el vector (if the N\'{e}el vector is along $a$ as in Fig.~\ref{fig:1}), and symmetry-allowed out-of-plane canting. As a result there can be a $\bm \sigma_{\rm AH}$ along the 2D plane normal.

We start by considering a three-term Hamiltonian:
\begin{eqnarray}\label{eq:Hnif22d}
H &=& H_t + H_{\rm ex} + H_{\rm so} \\\nonumber
&=& -t \sum_{\langle i\alpha,j\beta\rangle\gamma} c_{i\alpha\gamma}^\dag c_{j\beta\gamma} - J_{\rm ex} \sum_{i\alpha\gamma\delta}  (\hat{n}_{\alpha} \cdot \bm \sigma_{\gamma\delta} ) c_{i\alpha\gamma}^\dag c_{i\alpha\delta}\\\nonumber
&{}& + {\rm i}\lambda_{\rm so}\sum_{\langle i\alpha,j\beta\rangle\gamma\delta} [(\hat{\eta}_{i\alpha,j\beta}\times \hat{r}_{i\alpha,j\beta})\cdot \bm \sigma_{\gamma\delta}] c_{i\alpha\gamma}^\dag c_{j\beta\delta},
\end{eqnarray}
where $H_t$, $H_{\rm ex}$, and $H_{\rm so}$ capture nearest-neighbor spin-independent hopping, exchange coupling to a staggered in-plane on-site exchange field, and spin-orbit coupling, with the coupling parameters $t$, $J_{\rm ex}$, and $\lambda_{\rm so}$, respectively. Here $i,j$ label unit cell, $\alpha,\beta=A,B$ label sublattice, and $\gamma,\delta$ label spin. The directions of the local exchange fields $\hat{n}_{\alpha}$ and the spin-orbit coupling vector $\hat{\eta}_{i\alpha,j\beta}$ are represented respectively by in-plane and the out-of-plane arrows in the unit cell shown in Fig.~\ref{fig:2}.

The momentum space Hamiltonian after Fourier transform is
\begin{eqnarray}
H_{\bm k} &=& -2t (\cos k_x + \cos k_y) \tau_x \sigma_0 - \frac{J_{\rm ex}}{\sqrt{2}} \tau_z (\sigma_x - \sigma_y)\\\nonumber 
&&- 2\lambda_{\rm so} (\cos k_x \tau_y \sigma_y + \cos k_y \tau_y \sigma_x).
\end{eqnarray}
where $k_{x,y}$ are in units of $1/a$, $a$ being the nearest neighbor distance, $\bm \tau$ is the Pauli matrix vector in the sublattice pseudospin space, and $\sigma_0$ is the identity matrix in real spin space. When only $H_t$ is present, the energy spectrum has four-fold degeneracies at zero energy along the $k_x \pm k_y = \pm \pi$ and $k_x \pm k_y = \mp \pi$ Brillouin zone boundaries. $H_{\rm so}$ removes these degeneracies except at the time-reversal-invariant momenta $(\frac{\pi}{2},\zeta \frac{\pi}{2})$ [or equivalently $(\zeta \frac{\pi}{2}, \frac{\pi}{2})$] where $\zeta = \pm 1$. Near the time reversal invariant momenta
\begin{eqnarray}\label{eq:H2Dvalley}
H_{\bm k,\zeta} &=& 2 (t \tau_x \sigma_0 +\lambda_{\rm so} \tau_y \sigma_y) k_x\\\nonumber
&&+ 2 \zeta (t \tau_x \sigma_0 + \lambda_{\rm so} \tau_y \sigma_x) k_y - \frac{J_{\rm ex}}{\sqrt{2}} \tau_z (\sigma_x - \sigma_y),
\end{eqnarray}
where $k_x$ and $k_y$ are small momentum deviations. When $J_{\rm ex} = 0$, the low-energy Hamiltonian maps to anisotropic Dirac cones at two valleys distinguished by $\zeta$ with 
\begin{eqnarray}
\epsilon_{\bm k,\zeta} = \pm 2\sqrt{(\lambda_{\rm so}^2+t^2) k^2 + 2\zeta t k_x k_y}.
\end{eqnarray}
Finite $J_{\rm ex}$ opens a gap at each valley. Note that if the exchange fields are rotated in-plane by $\pi/2$, replacing $\sigma_x - \sigma_y$ in the last term of Eq.~\ref{eq:H2Dvalley} by $\sigma_x + \sigma_y$, the exchange coupling term commutes with $H_{\rm so}$ at $k_x + \zeta k_y = 0$ and does not open a gap. It follows that when $J_{\rm ex}$ is small the Chern number contribution from valley $\zeta$ is $C_{\zeta} = \zeta$, {\it i.e.} the model system becomes a valley Chern insulator but the total Chern number and $\sigma_{\rm AH}$ vanish. In order to make $C$ nonzero so that the system hosts QAHE, we need to add mass terms to the Hamiltonian which have different values at the two valleys. We now show that these can be supplied by a Rashba spin-orbit coupling combined with an out-of-plane canting of the exchange fields, which are readily available in the more realistic system explained in the next section.  

The Rashba-type spin-orbit coupling contribution ($H_{\rm R}$) to the Hamiltonian differs from $H_{\rm so}$ by replacing $\hat{\eta}_{i\alpha,j\beta}$ in Eq.~\ref{eq:Hnif22d} with a vector along $z$ (or $-z$). To lowest order in $\bm k$
\begin{eqnarray}
H_{\rm R,\zeta} = 2\lambda_{\rm R} \tau_x (\zeta \sigma_x - \sigma_y),
\end{eqnarray}
where $\lambda_{\rm R}$ is a coupling constant. Note that $H_{\rm R,\zeta}$ anticommutes (commutes) with the exchange coupling term when $\zeta = 1 (-1)$, and therefore should have different effects on the gaps opened by finite $J_{\rm ex}$ at the two valleys. $H_{\rm R,\zeta}$ does not, however, supply gaps by itself when $J_{\rm ex}=0$ since it commutes with the hopping terms in Eq.~\ref{eq:H2Dvalley} along $k_x = k_y$ lines. Thus when $\zeta = 1$, $H_{\rm R,\zeta}$ effectively shifts $\bm k$ but does not close the gap, while when $\zeta = -1$ it can close the gap but cannot reopen it. To open the gap at the $\zeta = -1$ valley again we simply need a small constant gap term that anticommutes with $H_{\rm ex}$, which can be supplied by an out-of-plane uniform exchange field due to canting:
\begin{eqnarray}
H_{\rm cant} = -J_z\tau_0 \sigma_z,
\end{eqnarray}
where $\tau_0$ is an identity matrix in sublattice space.

\begin{figure}[h]
	\begin{center}
		\includegraphics[width=1.6in]{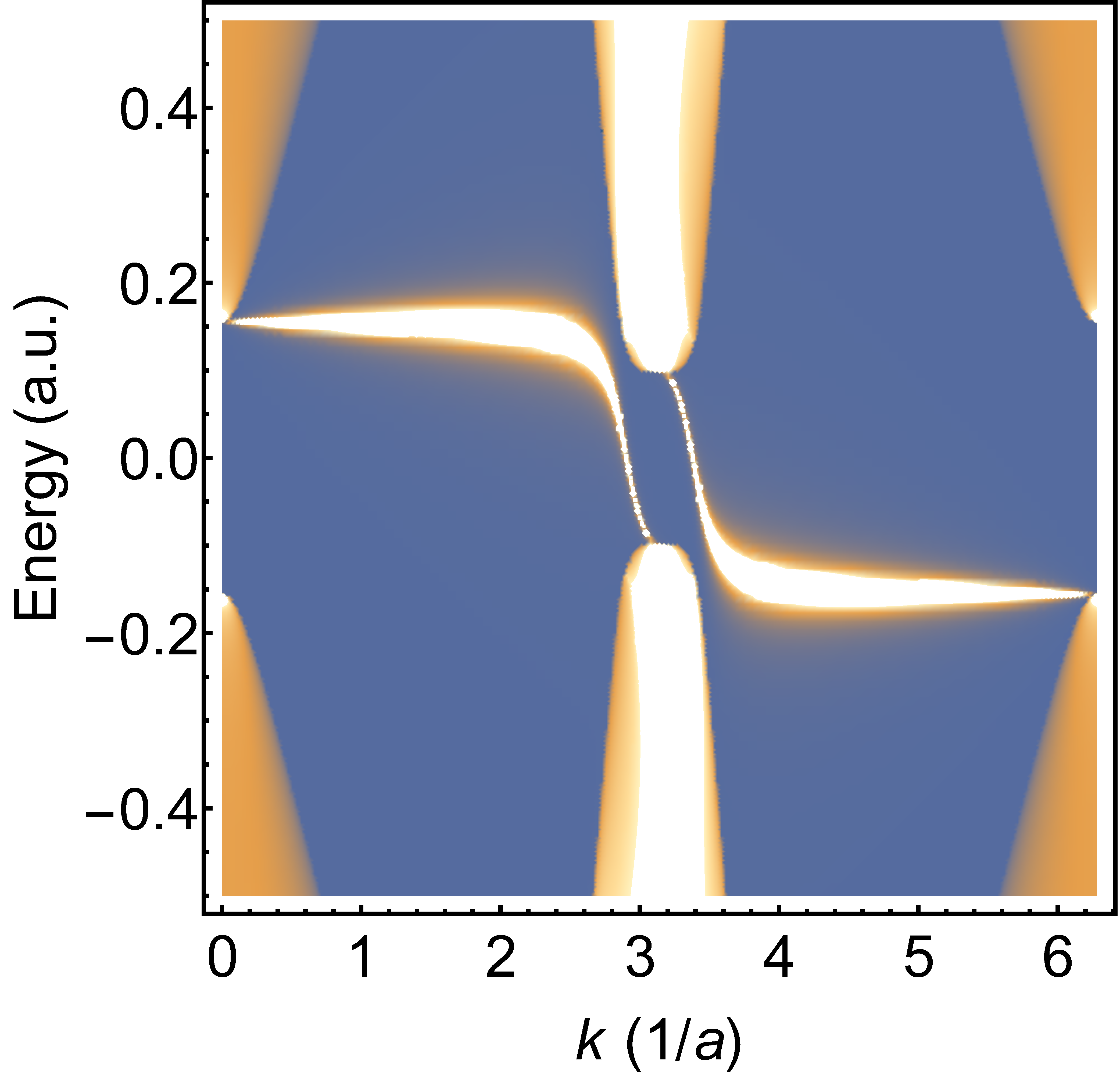}
		\includegraphics[width=1.6in]{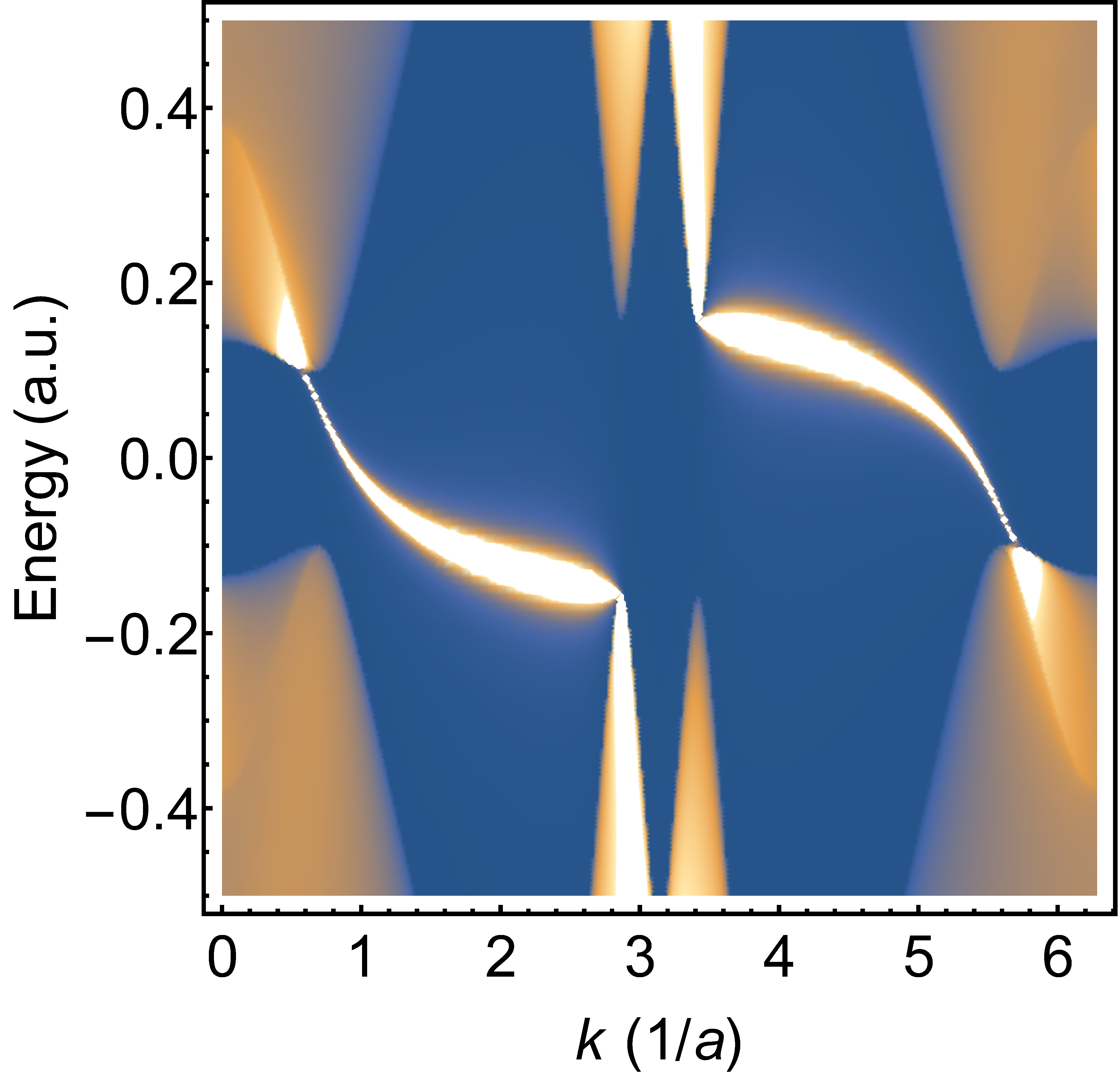}
	\end{center}
	\caption{\label{fig:nif2-2D-edge} (Color online) Plots of the imaginary part of the edge Green's function \cite{williams_1982,Lopez-Sancho_1985} of semi-infinite systems of the model in Fig.~\ref{fig:2} with boundaries perpendicular to $\hat{x}+\hat{y}$ (left) and $\hat{x}-\hat{y}$ (right). Both edges show two branches of edge states within the bulk gaps. The model parameters used for this calculation were $t=1$, $\lambda_{\rm so} = 0.5$, $J_{\rm ex}=0.2$, $\lambda_{\rm R} = 0.2$, and $J_z = 0.1$.} 
\end{figure}

When all five terms are present in the Hamiltonian we obtain a total valence band Chern number $C = 2$ over broad ranges of parameter space, when the relation $t_{\rm so}\gg J_{\rm ex}\sim \lambda_{\rm R}$ is qualitatively satisfied as discussed above. The corresponding edge states for a typical set of parameter values are illustrated in Fig.~\ref{fig:nif2-2D-edge}.

It is worth noting that the model above has an emergent particle-hole symmetry $\mathcal{C} \equiv \tau_z \sigma_y$, which makes it belong to the class C of the ten-fold Altland-Zirnbauer classification, and have a $2\mathbb{Z}$ topological invariant \cite{ryu_2010,chiu_2016}. This can be taken as one reason for the $C=2$. The $\mathcal{C}$ symmetry also leads to an interesting connection between the QAHE model and class C topological superconductors \cite{supp}. 

\begin{figure}
	\begin{center}
		\includegraphics[width=1.9 in]{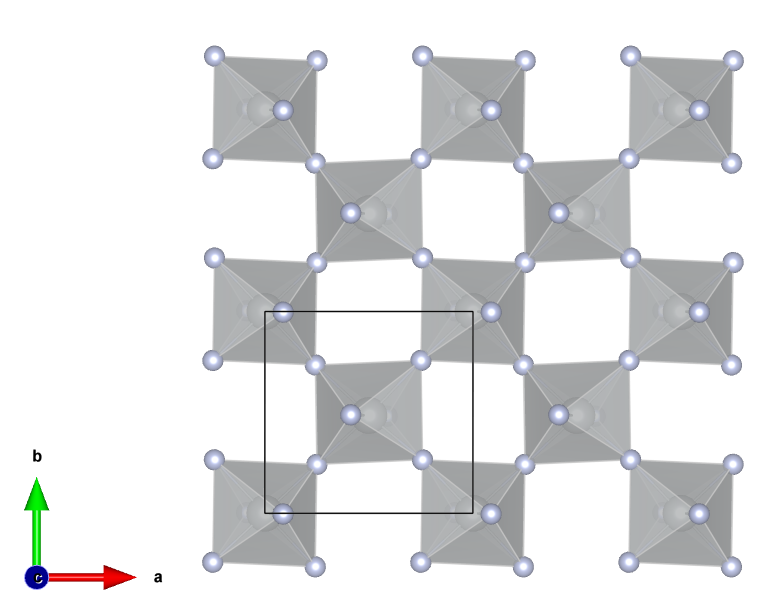}
	\end{center}
	\caption{Top view of a (001) layer of an orthorhombically distorted perovskite with the chemical formula ABX$_3$. Only the B and X atoms forming octahedra are shown.}\label{fig:orthofe-2D}
\end{figure}

\textit{QAHE in 2D Canted AFMs: Realistic Considerations---}
We now show that this toy model describes the low-energy physics of a realistic system. We first note that the (011) plane of NiF$_2$, including the neighboring F atoms, looks similar to the (001) plane of a perovskite with orthorhombic distortions, as shown in Fig.~\ref{fig:orthofe-2D}. Orthorhombic distortion is common in perovskites with the Goldschmidt tolerance factor smaller than 1 \cite{goodenough_1970}, and is indeed the symmetry reason for canting in magnetic perovskites such as orthoferrites \cite{bozorth_1958, dzyaloshinskii_1958}. In the classification of orthorhombic distortions due to Glazer \cite{glazer_1975}, the type of distortion in Fig.~\ref{fig:orthofe-2D} is referred to as $a^{-}b^{-}c^0$, where $a^-$ means rotation of octahedra around the $a$ axis of the so-called pseudocubic cell defined by the undistorted lattice with one formula unit per cell. Note that to maintain their bonding with the common oxygen atom (assuming the material under discussion is an oxide), two neighboring octahedra along the directions perpendicular to $a$ must rotate oppositely. The $-$ in $a^-$ means that the two neighboring octahedra along the direction of $a$ also rotate oppositely (or out-of-phase). $b^-$ has a similar meaning, and for simplicity we take the rotation angle identical to that with respect to $a$. $c^0$ means there is no rotation around the $c$ axis. We find that the rotation around $c$ does not lead to qualitatively different behaviors of the model detailed below. In reality, depending on which direction is defined as $c$ the orthorhombic distortion along the $a$ and $b$ axes can be either $a^{-}b^{-}$ or $a^{-}b^{+}$. The latter, however, makes the unit cell four time larger than the undistorted lattice. To make direct comparison with the toy model above we consider only $a^{-}b^{-}$, which only doubles the unit cell.

For a transition metal ion in an octahedral crystal field, its $d$ orbitals are splitted into $e_g$ and $t_{2g}$ spaces separated by an energy scale called $10Dq$ which is usually several eV. The $t_{2g}$ space in particular has an effective orbital angular momentum $\tilde{l} = 1$, with $d_{xy}$, $-\frac{1}{\sqrt{2}} (id_{zx}\pm d_{yz})$ corresponding to $\tilde{l}_z = 0,\pm 1$ eigenstates, respectively \cite{abragam_1970,jackeli_2009}. The atomic spin-orbit coupling further splits the $t_{2g}$ space into a two-fold effective total angular momentum $\tilde{j} = \frac{1}{2}$ space and a four-fold $\tilde{j} = \frac{3}{2}$ space. For $4d$ and $5d$ elements with relatively large spin-orbit coupling and partially-occupied $\tilde{j} = \frac{1}{2}$ space, one can describe the low-energy physics of the system by projecting the full Hamiltonian into this pseudospin space \cite{jackeli_2009}, with 
\begin{eqnarray}
&&|\tilde{\uparrow} \rangle =  (d_{yz,\downarrow}  + i d_{zx,\downarrow} + d_{xy,\uparrow})/\sqrt{3},\\\nonumber
&&|\tilde{\downarrow} \rangle =  (-d_{yz,\uparrow}  + i d_{zx,\uparrow}  + d_{xy,\downarrow})/\sqrt{3},
\end{eqnarray}
where we have ignored the influence of a possible tetragonal distortion \cite{jackeli_2009}, which we found to be of secondary effects on the main conclusion.

We consider first the effect of orthorhombic distortion on the hopping between neighboring $d$ orbitals in this pseudospin space. For an $a^-$ rotation by angle $\theta$, when $\theta$ is small, we found that the hopping along $a$ between the two sublattices through the intermediate oxygen atom, written as a $2\times 2$ matrix in the pseudospin space of each ion's local coordinate system, is 
\begin{eqnarray}
t_{AB,a} = t_{dd} (\sigma_0 + 2 i \theta \sigma_x),
\end{eqnarray}
where $t_{dd} = \frac{2V_{pd\pi}^2}{3(\epsilon_p - \epsilon_d)}$, but hopping along $b$ does not change up to the 1st order in $\theta$. In contrast, the $b^-$ rotation by $\theta$ leads to 
\begin{eqnarray}
t_{AB,b} = t_{dd} (\sigma_0 - 2 i \theta \sigma_y),
\end{eqnarray}
with the hopping along $a$ unchanged up to the 1st order in $\theta$. Therefore, the out-of-phase orthorhombic rotation leads to an spin-orbit coupling exactly the same as that in the model above (Eq.~\ref{eq:Hnif22d}), if one performs a $-\pi/2$ rotation around $z$ in the spin space of the latter.

Moreover, if we assume there to be a staggered in-plane Zeeman field coupled to the real spin in the model with the same directions as that in Fig.~\ref{fig:2}, in the pseudospin space such a term becomes
\begin{eqnarray}
H_{\rm ex} = -J_{\rm ex} \left [\frac{\sqrt{2}}{6} \tau_z (\sigma_x + \sigma_y) - \frac{\theta}{3}\tau_0\sigma_z \right ].
\end{eqnarray}
The first term is identical to that in Eq.~\ref{eq:Hnif22d} up to a $\pi/2$ rotation in spin space. In addition, the 2nd term, automatically given by the orthorhombic distortion, is exactly the out-of-plane canting $H_{\rm cant}$.

Finally, as discussed in \cite{khalsa_2013}, when the mirror symmetry with respect to the plane of the transition metal ions is broken, a Rashba-type spin-orbit coupling in the $t_{2g}$ space can arise from the new two-center $p-d$ integrals $E_{x,zx}$ along $y$ and $E_{y,yz}$ along $x$, that are allowed by symmetry. This can be understood as due to an uniform displacement of oxygen atoms that used to be coplanar with the metal ions to the out-of-plane direction. As a first approximation we assume the symmetry breaking leading to the Rashba spin-orbit coupling is independent of the orthorhombic distortion. After obtaining the effective $d-d$ hopping with this effect included, and projecting it to the pseudospin space, we obtain a Rashba-type spin-dependent hopping:
\begin{eqnarray}
t^{\rm R}_{AB,\pm a} = \pm i t_{\rm R} \sigma_y,\,\, t^{\rm R}_{AB,\pm b} = \pm i t_{\rm R} \sigma_x,
\end{eqnarray}
where $\pm $ means the hopping along opposite directions takes opposite signs. Different contributions to $t_{\rm R}$ can be found in \cite{khalsa_2013}. Although the Rashba-type hopping is different from that in the toy model, on can check that at the two valleys the (anti)commutation relations between the different terms discussed for the toy model still hold. The particle-hole symmetry $\mathcal{C}$ is also the same. Thus based on the discussion in the previous section, in large regions of the parameter space spanned by $\theta$, $J_{\rm ex}/t_{dd}$, and $t_{\rm R}/t_{dd}$ the present system can be a QAHE with $2\mathbb{Z}$ Chern number. For example, using $\theta = 11^\circ$, $J_{\rm ex} = 0.5 t_{dd}$, $t_{\rm R} = 0.1 t_{dd}$, which are reasonable in the $\tilde{j} = \frac{1}{2}$ space, we got a $C = 2$.

Since above model requires broken mirror symmetry, a practical structure based on it should be an interface of orthorhombically distorted $4d$ or $5d$ perovskite which is insulating in the bulk. Note that many bulk $5d$ oxides derived from the perovskite structure already have canted AFM orders, such as the spin-orbit Mott insulator Sr$_2$IrO$_4$ \cite{cao_1998, kim_2008, kim_2009, jackeli_2009} and the post perovskite CaIrO$_3$ \cite{cheng_2011}. If not one can use another canted insulating AFM with similar structure, e.g., orthoferrite, to induce the staggered exchange field at the interface. It is also possible to consider a single layer of the structure shown in Fig.~\ref{fig:orthofe-2D} sandwiched between two wide-gap insulators, with one of them a canted AFM. Ideally the electronic structure at the interface should have the correct filling with one electron (or hole) in the $\tilde{j} = \frac{1}{2}$ subspace of $t_{2g}$. The advantage of specializing to the $\tilde{j} = \frac{1}{2}$ space is that all the parameters mentioned above are on the same order of magnitude, which makes it easy for the system to be tuned into a QAHE state. 


Note added: In the preparation of this paper we noticed a related preprint \cite{smejkal_2019} on the AHE in collinear antiferromagnets.

\begin{acknowledgments}
The authors are grateful to Satoru Nakatsuji, Carl Patton, Jianshi Zhou, and Kate Ross for insightful comments and suggestions. HC also thanks the hospitality of the Institute for Solid State Physics at the University of Tokyo during his visit in 2018, when his contribution to this work was done. AHM was supported by SHINES. HC was supported by the start-up funding from CSU.
\end{acknowledgments}

\bibliography{ref}

\end{document}